\documentclass[%
 aip,
 amsmath,amssymb,
 reprint,%
]{revtex4-1}

\usepackage{graphicx}% Include figure files
\usepackage{setspace}
\usepackage{xspace}
\usepackage{dcolumn}% Align table columns on decimal point
\usepackage{bm}
\usepackage{xcolor}
\usepackage{float}% bold math
\usepackage[utf8]{inputenc}
\usepackage[T1]{fontenc}
\usepackage{mathptmx}
\usepackage{etoolbox}

\makeatletter
\def\@email#1#2{%
 \endgroup
 \patchcmd{\titleblock@produce}
  {\frontmatter@RRAPformat}
  {\frontmatter@RRAPformat{\produce@RRAP{*#1\href{mailto:#2}{#2}}}\frontmatter@RRAPformat}
  {}{}
}%
\makeatother
\begin{document}

\title{Stress Accommodation in Nanoscale Dolan Bridges Designed for Superconducting Qubits}

\author{S. Skinner-Ramos}
 \affiliation{Sandia National Laboratories, Albuquerque, New Mexico 87185, USA}
 \author{M. L. Freeman}%
\affiliation{Sandia National Laboratories, Albuquerque, New Mexico 87185, USA}

\author{D. Pete}%
\affiliation{Center for Integrated Nanotechnologies, Sandia National Laboratories, Albuquerque, New Mexico 87185, USA}
\author{R. M. Lewis}
\affiliation{Sandia National Laboratories, Albuquerque, New Mexico 87185, USA}
\author{M. Eichenfield}
 \affiliation {Sandia National Laboratories, Albuquerque, New Mexico 87185, USA}
 \affiliation{Wyant College of Optical Sciences, University of Arizona, Tucson, Arizona 85721, USA}
\author{C. Thomas Harris$^{\ast,}$}
\affiliation {Center for Integrated Nanotechnologies, Sandia National Laboratories, Albuquerque, New Mexico 87185, USA}

 \email{ctharri@sandia.gov}

\title{Stress Accommodation in Nanoscale Dolan Bridges Designed for Superconducting Qubits}

\begin{abstract}
 Josephson junctions are the principal circuit element in numerous superconducting quantum information devices and can be readily integrated into large-scale electronics.  However, device integration at the wafer scale necessarily depends on having a reliable, high-fidelity, and high-yield fabrication method for creating Josephson junctions.  When creating Al/AlO$_{\rm{x}}$ based superconducting qubits, the standard Josephson junction fabrication method relies on a sub-micron suspended resist bridge, known as a Dolan bridge, which tends to be particularly fragile and can often times fracture during the resist development process, ultimately resulting in device failure. In this work, we demonstrate a unique Josephson junction lithography mask design that incorporates stress-relief channels. Our simulation results show that the addition of stress-relief channels reduces the lateral stress in the Dolan bridge by more than 70\% for all the bridge geometries investigated. In practice, our novel mask design significantly increased the survivability of the bridge during device processing, resulting in 100\% yield for over 100 Josephson junctions fabricated.

\textbf{Keywords:} Josephson junctions, superconducting qubits, stress modeling.

\end{abstract}

\maketitle

\section{Introduction}

Owing to their nonlinear inductance properties, Josephson junctions (JJs) are central to many quantum information (QI) systems,~\cite{Minev2021,Clarke} such as superconducting quantum interference devices (SQUIDS)~\cite{Jaklevic1964} and superconducting qubits.~\cite{Makhlin1999,Osman2021} The standard JJs used in superconducting qubits are based on the Al/AlO$_{\rm{x}}$ material system.  When viewed in cross-section, these Al/AlO$_{\rm{x}}$ JJs are comprised of two overlapping Al superconductors that are separated by a thin AlO$_{\rm{x}}$ tunneling barrier. When coupled with a capacitor, these junctions, either solitary or paired to fashion a SQUID, form an anharmonic oscillator with broken energy level degeneracy that enables the creation of a quasi two-level system, the transmon \cite{koch2007}. A standard, self-aligned JJ lithography mask consists of two perpendicularly oriented fingers (Fig.\,1(a) (inset)) that, when patterned on a bilayer resist stack, result in the formation of a very narrow Dolan bridge~\cite{Dolan1977, Bilmes_2021}. Because the Dolan bridge is fragile and notoriously prone to fracture ~\cite{Slichter} as shown in Fig.\,1(a), the challenge is to fabricate JJs reliably at scale. The most common state-of-the-art fabrication methods that help to preserve the integrity of the Dolan Bridge rely on cold development techniques~\cite{Kreikebaum_2020,Kirill2011,Ocola2004,Wenchuang2004} with a low ultrasonication power~\cite{Kreikebaum_2020}, the use of orthogonal resists~\cite{Cord2006,Tanner2008}, and techniques that go beyond the use of Dolan Bridges, such as the Manhattan style ~\cite{Potts2001,Kreikebaum_2020,SergioO2012} and the bridge-free methods~\cite{Lecocq_2011, Shankar_2013, Minev_2019}. However, these more reliable fabrication methods are also fairly cumbersome and somewhat time-consuming to implement compared to using more commonplace resists like poly(methyl methacrylate) (PMMA) and methyl methacrylate (MMA) at room temperature. 
Mechanically stress-induced fracturing of the Dolan bridge is thought to be the primary failure mechanism for standard, self-aligned lithography masks when using PMMA and MMA resists. To address this issue, we augment a self-aligned mask to include stress-relief channels on both sides of the Dolan bridge, which reduces stress locally.  Our innovative JJ mask design allows for a simple and efficient fabrication approach to reliably produce JJs with intact Dolan bridges at large scales. To demonstrate the reliability of our mask design, we fabricated over 100 JJs with $\sim$100\,nm critical feature sizes that exhibited no fractures. We performed finite-element simulations to examine the mechanical behavior of  JJ resist masks and to quantify the reduction in mechanical loading when using stress-relief channels. To characterize the performance of our stress relieved JJ fabrication in a typical QI device, we fabricated a transmon qubit using the stress-relief lithography mask design we discuss here and measured its properties at dilution refrigerator temperatures.

  %%%%% Fig. 1 %%%%%
\begin{figure*}[t]
	\centering
	\includegraphics[scale=0.57]{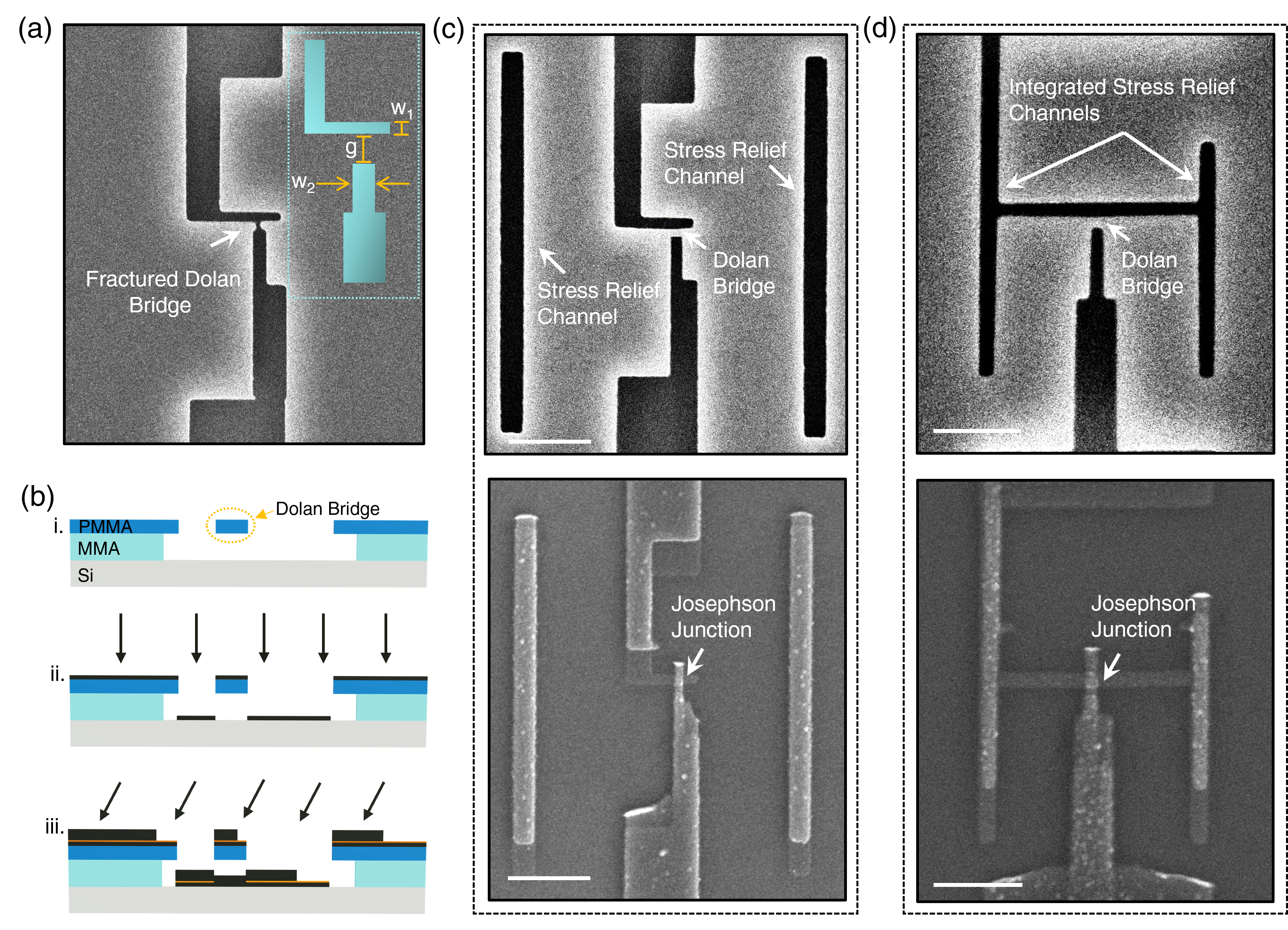}
	\caption{(a) Scanning electron micrograph (SEM) image of the resist mask of a self-aligned JJ with a fractured Dolan bridge, and (inset) the design parameters that set the geometry of the bridge $w_1$, $w_2$ and $g$.(b) Fabrication process flow for JJs using the shadow evaporation technique. The dark arrows illustrate the angle of the metal evaporation direction with respect to the sample. (c) SEM image of the resist mask after adding stress-relief channels (top) and after the metal liftoff process (bottom). (d) SEM image of the resist mask with integrated stress-relief channels (top) and after the metal liftoff process (bottom). All scale bars are 1\,$\mu$m.}
\end{figure*}
%%%%%%%%%%    

\section{Device Fabrication}

Our JJs were fabricated using a double-angle evaporation technique, the most common JJ fabrication process, to create an Al/AlO$_{\rm{x}}$/Al material stack.~\cite{Dolan1977,Kreikebaum_2020,Pop2012,Potts2001} The fabrication process starts by spin-coating a bilayer resist stack comprised of a 760\,nm-thick MMA EL 13 film and a 180\,nm-thick 950 PMMA A3 layer on top of a Si substrate. We then patterned the resist using electron beam lithography at an acceleration voltage of 30 keV and developed the resist in a methyl isobutyl ketone:isopropyl alcohol (MIBK:IPA) solution, which selectively undercuts the MMA, leaving behind a thin, suspended PMMA layer (the Dolan bridge), as seen in Fig.\,1(b,\,i). To remove any PMMA residue from the Si substrate, the sample underwent an oxygen plasma etch (100\,W, 2\,min.), and immediately thereafter it was loaded into an e-beam evaporator for metallization.  After pumping to a base pressure of 1x$10^{-7}$\,Torr, a 20\,nm-thick layer of Al was deposited normal to the substrate surface (Fig.\,1(b,\,ii)). Immediately after evaporation, the chamber was flooded with O$_{2}$ to create a static O$_{2}$ background at a pressure of 500\,mTorr for 10 minutes to oxidize the Al, forming a thin ($\sim$10\AA) layer of AlO$_{\rm{x}}$ that serves as the JJ tunnel barrier. Finally, after quickly returning to a pressure of 1x$10^{-7}$\,Torr, a 40\,~nm-thick Al layer was deposited at an angle of 30$^{\circ}$ to normal (Fig.\,1(b,\,iii)).  The sample  was then transferred to an acetone bath for metal liftoff. This double-angle evaporation technique, or shadow evaporation technique, produces metal features that are identical to the resist mask and are offset laterally such that their overlap under the shadow of the Dolan bridge creates a JJ. 
A fully completed JJ employing strain relief channels is shown in Fig.\,1(c) (bottom), and the bilayer resist mask pattern used to produce this device is shown in Fig.\,1(c) (top).  As shown by the bilayer resist mask in Fig.\,1(a), the Dolan bridge completely fractures in the absence of strain-relief channels.  Because the (electrically floating) metal islands that form from the strain relief mask result in a capacitance near the JJ (estimated to be $\sim$1\,fF), we developed a second design that integrates the strain-relief channels into the JJ electrodes, as shown in Fig.\,1(d).  This integrated strain-relief design electrically clamps the potential of the strain-relief channels to that of the top electrode, removing the risk of having unknown potentials on nearby metallic islands (Fig.\,1(c) (bottom)).

 %%%%% Fig. 2 %%%%%
\begin{figure*}[t]
\centering
	\includegraphics[scale=0.5]{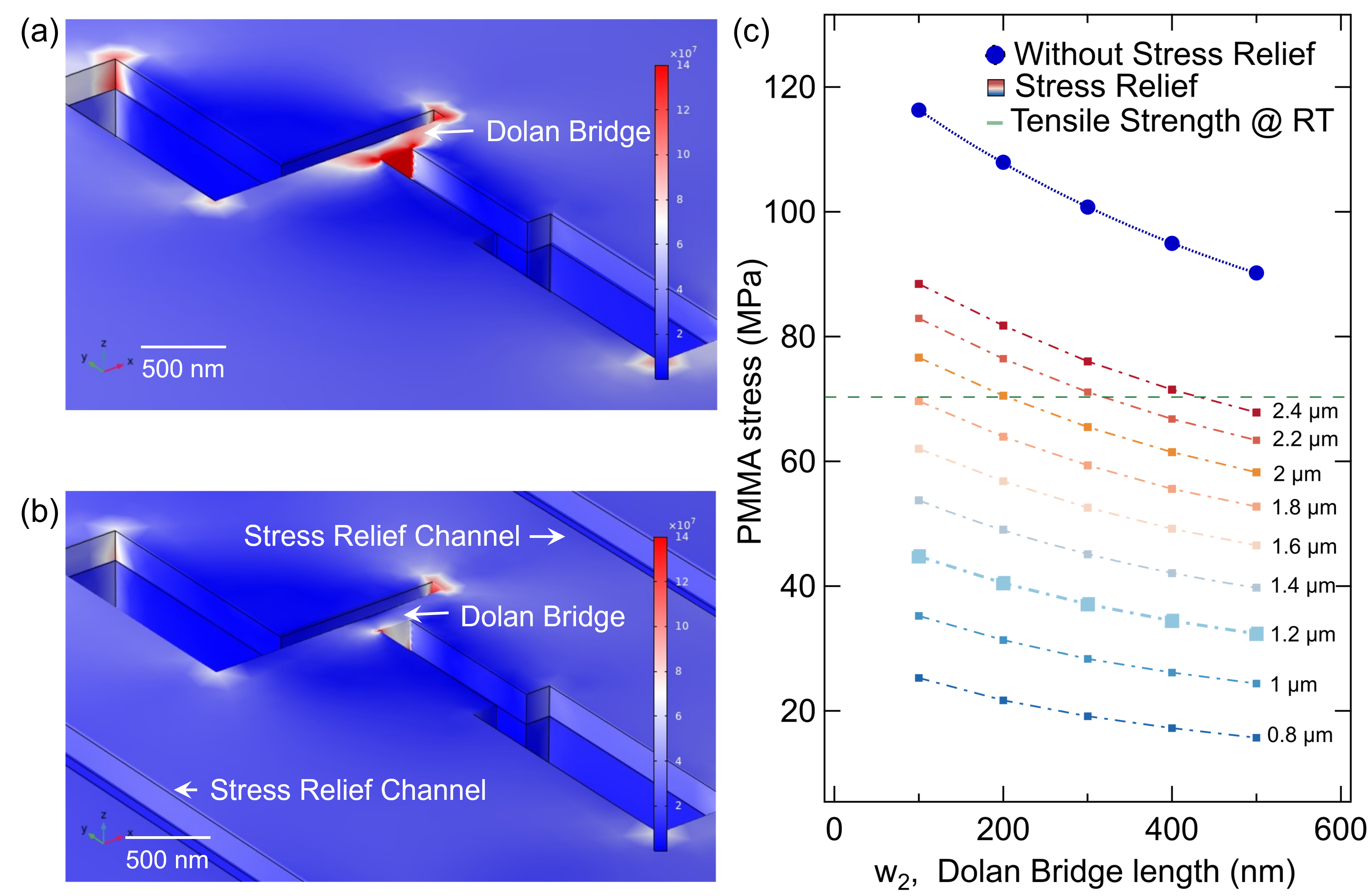}
	\caption{Simulation results of stress in the Dolan Bridge after resist stack development (a) without the stress-relief channels and (b) with stress-relief channels. The color bars represent stress in units of Pa. (c)  Calculated average value of the lateral stress-tensor components from the finite element simulation of the Dolan bridge without stress-relief channels (blue circles) and with stress-relief channels included (gradient with squares). Each trend corresponds to an equidistant spacing of the stress relief channels from the Dolan bridge, 0.8 $\mu$m to 2.4 $\mu$m.$w_1$ and $g$ are both kept constant at 100\,nm.
}
     
\end{figure*}
%%%%%
	
\section{Finite Element Analysis}
    
We performed finite element method simulations of the Dolan bridge domain for a standard, self-aligned JJ \textit{with} and \textit{without} stress relief channels (Fig.\,2(a) and (b)) using COMSOL Multiphysics software. For the simulation, we used the actual device dimensions and bilayer resist stack material properties. We measured the material stack’s induced stress due to the resist deposition on the substrate independently for both MMA ($\sim$13\,MPa) and PMMA ($\sim$44\,MPa) layers using a Tencor Flexus tool. The critical geometries affecting Dolan bridge yield and JJ area are the width of the horizontal finger $w_1$, the width of the vertical finger $w_2$, and the gap between them $g$. 

For the fabricated devices and the finite-element model, $w_1$ and $g$ are both kept constant at 100\,nm to maintain the same evaporation angle, while $w_2$ (Dolan bridge length) varied from 100\,nm to 500\,nm in increments of 100\,nm. We calculated the intrinsic lateral stress-tensor components in the Dolan bridge for vertical finger widths $w_2$ and found them in excess of the maximum tensile strength of PMMA at room temperature($\sim70$\,MPa).~\cite{LIU2008,JeffreyPmma2015,AbdelWAHAB2017}.
To examine the effects of stress-relief channels on the Dolan bridge's point of failure, we repeated the FEM simulations, this time including channels (0.25\,$\mu$m wide, 5\,$\mu$m long) equidistant from the Dolan bridge, identical to those fabricated in Fig.\, 1(c), (Fig.\,2(b)). A parameter sweep varying the distance between the stress relief channel and the Dolan bridge, from 0.8 $\mu$m to 2.4 $\mu$m, for each $w_2$ was performed. 
    %%%%% Fig. 3 %%%%%
\begin{figure}[h]
        \centering
	\includegraphics[scale=0.6]{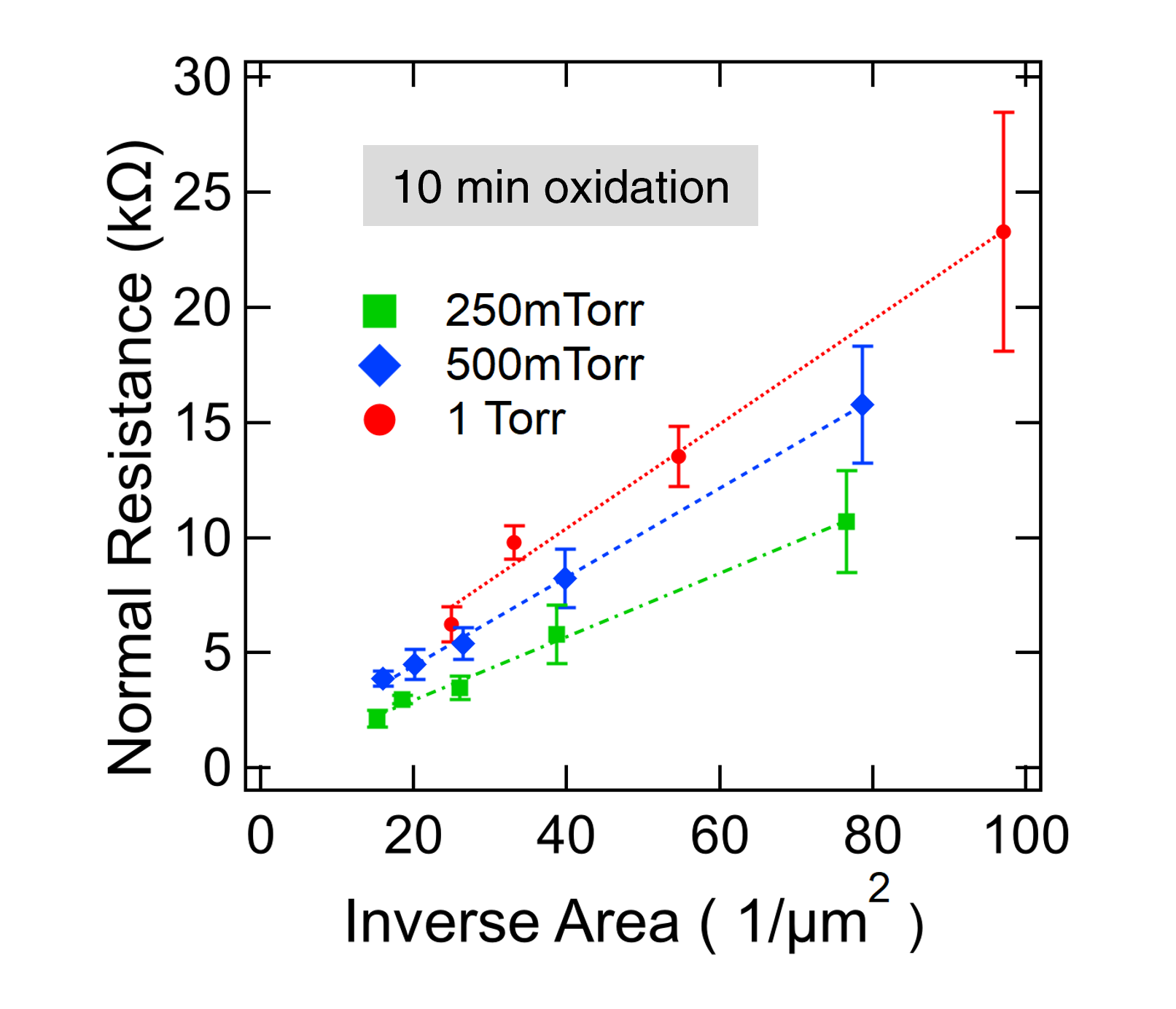}
	\caption{ Normal resistance values of JJs with varying junction areas fabricated using three oxidation pressures: 1\,Torr, 500\,mTorr and 250\,mTorr. The areas fabricated for the Josephson junctions were 100\,nm\,x\,100\,nm, 100\,nm\,x\,200\,nm, 100\,nm\,x\,300\,nm, 100\,nm\,x\,400\,nm, and 100\,nm\,x\,500\,nm.}
   
\end{figure}
The simulation results shown in Fig.\,2(c) reveal that when the stress-relief channels are separated from the Dolan bridge around a distance of 2 $\mu$m or greater, the induced lateral stress approaches the tensile strength of PMMA, which indicates that these geometries will likely result in a collapsed bridge at development. 
For the actual device, a channel distance of 1.2 $\mu$m, corresponding to a resulting induced stress that is ~60\% of the PMMA tensile strength, was chosen. With the stress-relief channels at this particular position, the intrinsic stress for all vertical finger widths in the Dolan bridge is approximately 30\% of the calculated stress for the JJ design without the stress relief channels. 
Our observations of JJ fabrication reinforce our simulation results. Dolan bridges fabricated using a self-aligned lithography mask with a PMMA/MMA resist stack developed at room temperature are heavily prone to fracture and fail nearly 100\% of the time during the development process. Based on our simulation results, we ascribe these failures to the stress within the Dolan bridge surpassing the tensile strength of the PMMA. 
By way of comparison, we observed 100\% yield in over 100 JJs fabricated using our mask design with integrated stress relief channels.

 %%%%% Fig. 4 %%%%%
\begin{figure*}[t]
	\begin{centering}
	\includegraphics[scale=0.65]{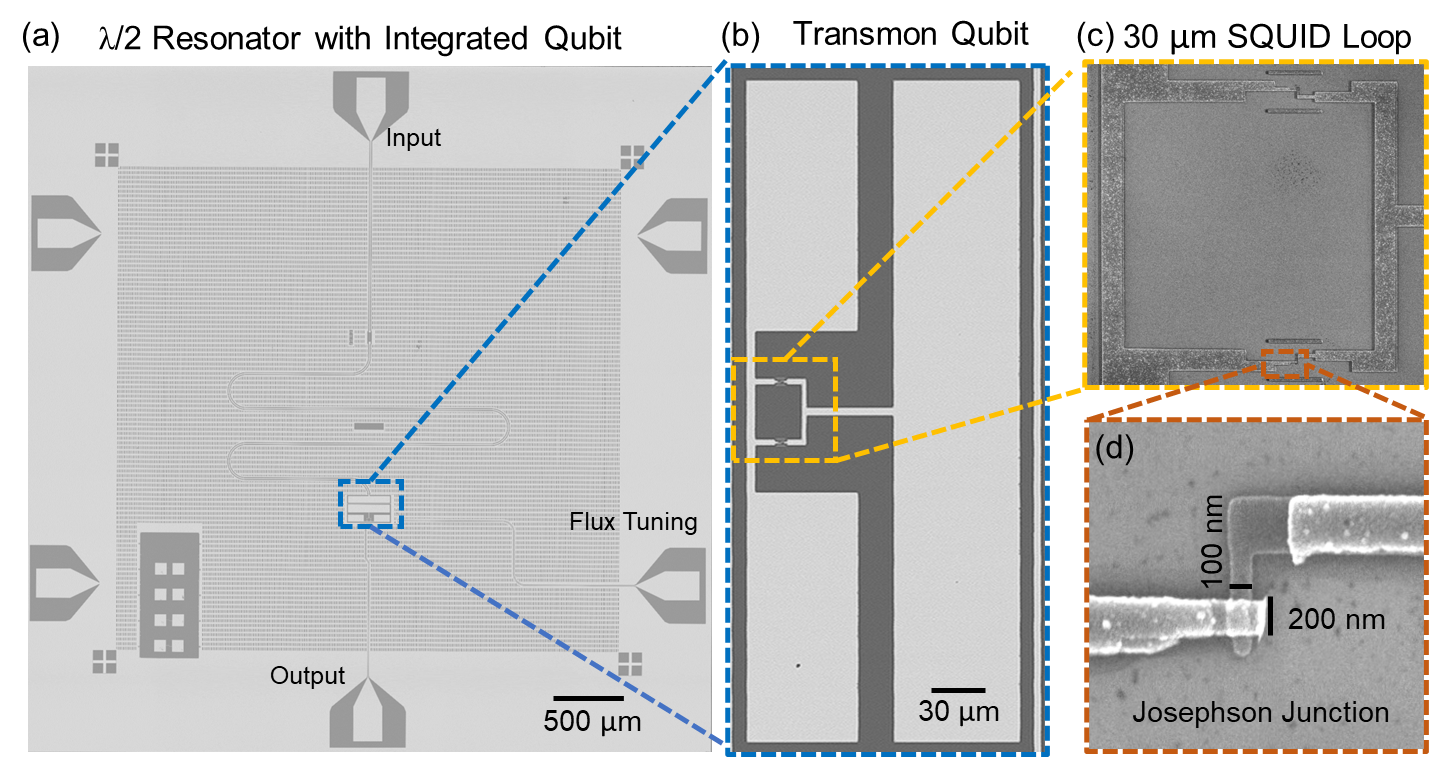}
   \caption{ (a) Optical image of the readout resonator, transmon qubit, and flux tuning line after the fabrication process. The resonator is a $\lambda$/2 coplanar waveguide that couples to the qubit. (b) Optical image of the entire transmon qubit consisting of an interdigitated capacitor and a 30\,$\mu$m SQUID loop. SEM image taken of (c) the SQUID loop, and (d) a Josephson junction with an area of $100~\rm{nm}~x~200~\rm{nm}$.
}
    \end{centering}
\end{figure*}
    %%%%%%%%%%   

\section{Josephson Junction Characterization}

Key physical parameters of the JJ, such as  its critical current $I_c$ and its capacitance $C_J$, may be tailored by appropriately adjusting the junction overlap geometry and tunnel barrier thickness. These foregoing parameters are coupled directly to the Josephson energy $E_J$ and to the charging energy $E_c$ by \(E_J= \hbar I_c/2e\) and $E_c= e^2/2C_\Sigma$,~\cite{Clarke,Devoret1990, Berthold2016} where $C_\Sigma$ is the total capacitance shunting the JJ including $C_J$, $\hbar$ is the reduced Planck's constant, and $e$ is the electron charge.  The anharmonicity and charge dispersion of a transmon qubit are determined by the ratio of $E_J$ and  $E_c$, while the level spacing is proportional to the square root of their product,~\cite{koch2007} and thus, well-controlled JJ fabrication is essential to accurately targeting qubit frequencies.  

Because the oxidation process can be used to modify the JJ's tunnel barrier thickness and to adjust its critical current, we evaluated the tunability of the JJs developed in this fabrication process by creating three sets of junctions fabricated at different oxidation pressures: 250\,mTorr, 500\,mTorr, and 1\,Torr. We electrically characterized these devices at room temperature, measuring their normal resistance $R_n$ , where each JJ's $I_c$ was calculated from $R_n$ using the Ambegaokar-Baratoff equation:~\cite{Baratoff1963} 

\begin{equation}
  I_c = \frac{\pi\Delta}{2e R_n }\,,
\end{equation} 
where $\Delta$ is the superconducting gap (182\,$\mu$eV for Al). 
The  results for $R_n$ are plotted in Fig.\,3.
Sample populations of ten or more JJs were measured electrically for each of the five nominal JJ areas at each of the three oxidation pressures tested. 
Error bars were generated from the standard deviation in these measurements.  As seen in Fig.\,3, a significant variation exists in the $R_n$ values for junctions with areas smaller than 0.025$\mu$m$^2$. 
This increase in resistance measurement error for smaller junctions is likely due to the native oxide that forms on the total exposed region of the junction, which for smaller junctions constitutes a greater fraction of the total oxidized region.
Furthermore, the oxidation step in our process is performed manually, which could also add uncertainty to the JJ's normal resistance.  
The critical currents for these sets of junctions were calculated using  Eqn.~\,1, resulting in three different $I_c$ ranges: 26-130\,nA (250\,mTorr), 18-73\,nA (500\,mTorr), and 11-45\,nA (1\,Torr).

\section{Transmon Qubit Measurements}

To demonstrate the utility of our JJs with the stress relief channels, we fabricated an Al/AlO$_{\rm{x}}$ transmon qubit~\cite{koch2007} using our JJ mask design and characterized its performance at dilution refrigerator temperatures ($T \approx 20$\,mK).
We employ an asymmetrical qubit crafted from two JJs with approximate critical currents of 9\,nA and 24\,nA and nominal areas of 150\,nm x 100 \,nm and 200\,nm x 100\,nm,  incorporated into a 30\,$\mu$m diameter SQUID loop with a  shunt capacitance of 67\,fF.
This design gave the qubit two sweet spots (i.e., $df_{01}/d\Phi=0$, where $f_{01}$ is the qubit transition frequency and $\Phi$ is magnetic flux) at 5.8\,GHz and 3.8\,GHz.
For state readout, we dispersively coupled the qubit to an Al $\lambda / 2$ superconducting resonator with a measured resonance frequency of 6.42\,GHz and a linewidth of 5.5\,MHz. The nominal qubit-resonator coupling, $g_{qr}/2 \pi \hbar$,  was 140\,MHz.  The fabricated transmon qubit and $\lambda / 2$ resonator are shown in Fig.\,4. We note that the dispersive readout chain included a TWPA amplifier \cite{Macklin} fabricated at MIT's Lincoln Laboratory.
Spectroscopy revealed that the qubit has a charging energy $E_c/2 \pi \hbar$ of 287\,MHz, corresponding to a shunt capacitance of approximately 67.4\,fF and a Josephson energy $E_J/h$ of 17.5\,GHz, due to the combined $I_c$ of the two JJs.

  %%%%% Fig. 5 %%%%%	
\begin{figure}[htbp]
	\centering
	\includegraphics[scale=0.74]{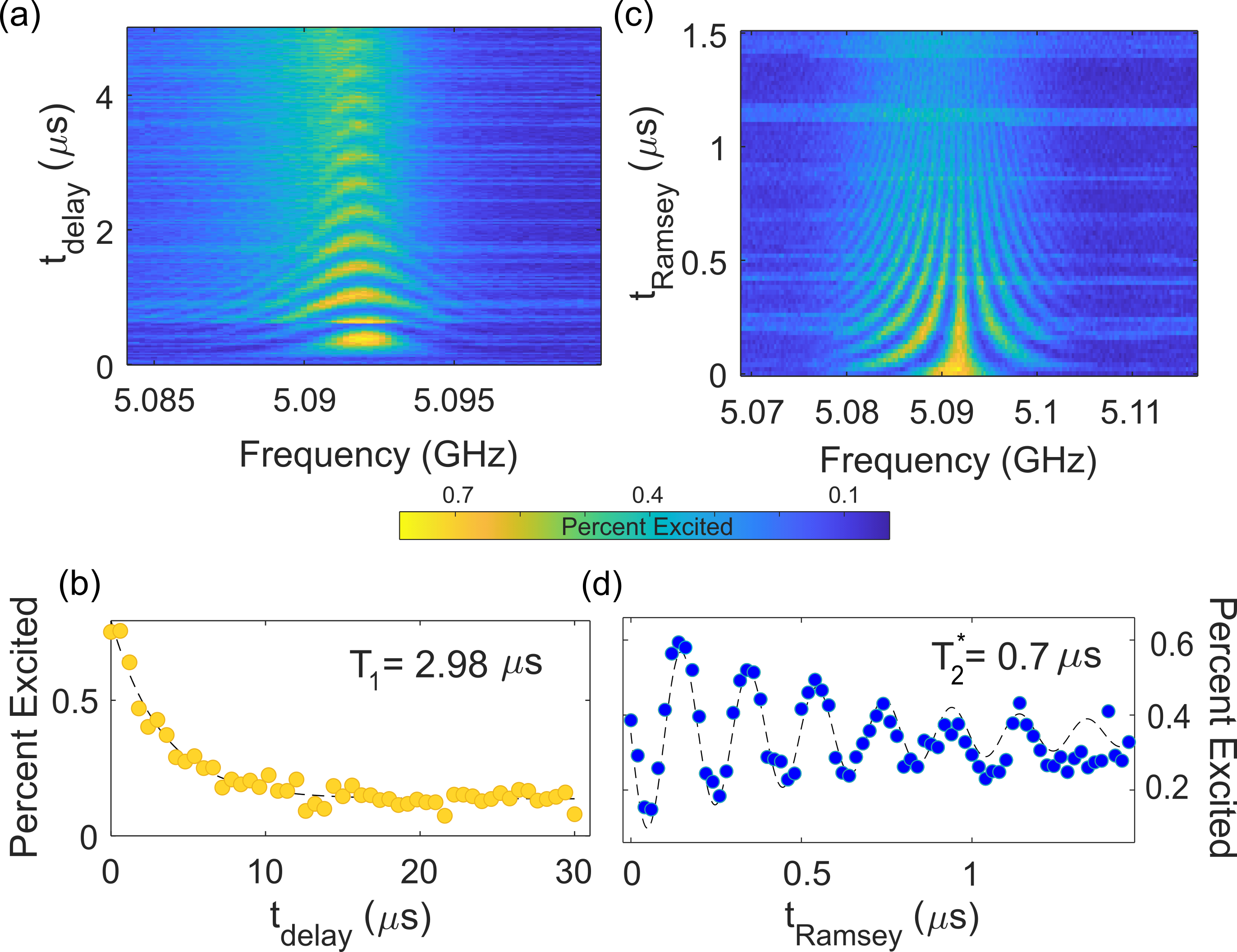}
	\caption{Qubit characterization. (a) Rabi oscillations of the qubit as a function of drive pulse length and frequency.  (b) Qubit lifetime $T_{1}$ versus the measured time delay. (c) Ramsey oscillations of the qubit to (d) characterize the qubit's coherence time ($T_{2}$). The color bars of (b) and (c) represent the readout amplitude and map to the qubit state where yellow represents $|0\rangle$ (ground state) and blue represents $|1\rangle$ (excited state).
}
\end{figure}
%%%%%%%%%%%    

The operating frequency of the qubit can be tuned by threading magnetic flux through the SQUID loop via a flux tuning line (see Fig\,4(a)). We characterized the qubit at a frequency of 5.09\,GHz to detune it away from the readout resonator and optimize readout fidelity. For qubit readout we used pulses of order 1\,$\mu$s. The $\pi$ and $\pi/2$ pulses were calibrated by adjusting the duration and frequency of the qubit drive as shown in Fig.\,5(a), with a $\pi$ pulse taking approximately 360\,ns. The qubit's lifetime $T_{1}$ was measured by exciting the qubit with a $\pi$ pulse and adjusting the delay time, $t_{\rm{delay}}$, before readout (Fig.\,5(b)).  
From this measurement we obtained a $T_{1}$ of 2.98\,$\mu$s.
To estimate the effect of the stress relief features on $T_1$, we performed finite element simulations of the electric field distribution in and around the qubit to approximate the participation ratio, $p_i$ ~\cite{Martinis}. We found $p_i \approx 0.001$ and thus they do not contribute significantly to the $T_{1}$ decay.

We acquired the qubit's coherence time $T_{2}^*$ by performing a Ramsey sequence of two $\pi/2$ pulses separated by a delay, $t_{\rm{Ramsey}}$, and followed by qubit readout.  Scanning the drive frequency at each time delay reveals Ramsey interference fringes (Fig.\,5(c)). A line cut through the Ramsey interference pattern detuned 5\,MHz from resonance displays oscillations and allows fitting to extract $T_{2}^*$, which we determine to be 0.7\,$\mu$s (Fig.\,5(d)). We note that the coherence time of the qubit is comparatively short at this bias point because of the large detuning from the optimal bias points at the qubit sweet spots and hence large $df_{01}/d\Phi$. The choice of this bias point was a compromise between readout fidelity and coherence.

\section{Conclusions}
   
In summary, we demonstrate a new, robust, and simple lithographic mask design for self-aligned JJs that incorporates stress relief channels.  
Our fabrication process does not require any additional steps beyond what is used in typical PMMA/MMA processing. 
Using our design, we successfully fabricated Dolan bridges that have critical dimensions of 100\,nm, and we observed a device yield of 100\% when fabricating over 100 JJs, which suggests this fabrication process is well suited for large-scale arrays of JJs.
Using finite element simulations, we determined that the intrinsic stress in nanoscale Dolan bridges fabricated without stress-relief channels is greater than the tensile strength of PMMA and is therefore likely to be the main contributor of bridge fracture. 
Our simulations also show that the stress in the Dolan bridge can be reduced by more than a factor of three, to values safely below the PMMA tensile strength, if stress-relief channels are incorporated in the JJ mask design. 
To demonstrate functionality of the JJs manufactured using our stress-relief design, we fabricated a transmon qubit coupled to a $\lambda/2$ resonator and show qubit operations, measuring the qubit's lifetime and coherence time, $T_{1}$ and $T_{2}^*$.

\section*{Acknowledgment}

The authors thank John Nogan and the CINT Integration Laboratory staff for process development assistance, Dr. Ethan A. Scott for reviewing this manuscript, and Prof. Rob Schoelkopf for donating the dilution fridge used. We acknowledge the use of a TWPA amplifier fabricated at MIT-LL under IARPA programs. This material is based upon work supported by the U.S. Department of Energy, Office of Science, National Quantum Information Science Research Centers, Quantum Systems Accelerator.
This work was performed, in part, at the Center for Integrated Nanotechnologies, a U.S. DOE, Office of Basic Energy Sciences, user facility. Sandia National Laboratories is a multimission laboratory managed and operated by National Technology \& Engineering Solutions of Sandia, LLC, a wholly owned subsidiary of Honeywell International, Inc., for the U.S. DOE's National Nuclear Security Administration under Contract No. DE-NA-0003525. The views expressed in the article do not necessarily represent the views of the U.S. DOE or the United States Government.

\bibliographystyle{elsarticle-num}
\section*{References}
\vspace{-4mm}
\bibliography{natbib}

\begin{thebibliography}{10}
\expandafter\ifx\csname url\endcsname\relax
  \def\url#1{\texttt{#1}}\fi
\expandafter\ifx\csname urlprefix\endcsname\relax\def\urlprefix{URL }\fi
\expandafter\ifx\csname href\endcsname\relax
  \def\href#1#2{#2} \def\path#1{#1}\fi

\bibitem{Minev2021}
Z.~K. Minev, Z.~Leghtas, S.~O. Mundhada, L.~Christakis, I.~M. Pop, M.~H. Devoret, Energy-participation quantization of josephson circuits, npj Quantum Information 7~(1) (2021).
\newblock \href {https://doi.org/10.1038/s41534-021-00461-8} {\path{doi:10.1038/s41534-021-00461-8}}.

\bibitem{Clarke}
J.~Clarke, F.~K. Wilhelm, Superconducting quantum bits, Nature 453~(1031) (2008).
\newblock \href {https://doi.org/10.1038/nature07128} {\path{doi:10.1038/nature07128}}.

\bibitem{Jaklevic1964}
R.~C. Jaklevic, J.~Lambe, A.~H. Silver, J.~E. Mercereau, Quantum interference effects in josephson tunneling, Phys. Rev. Lett. 12 (1964) 159--160.
\newblock \href {https://doi.org/10.1103/PhysRevLett.12.159} {\path{doi:10.1103/PhysRevLett.12.159}}.

\bibitem{Makhlin1999}
Y.~Mahklin, A.~Shnirman, Josephson-junction qubits with controlled couplings, Nature 398 (1999) 305--307.
\newblock \href {https://doi.org/10.1038/18613} {\path{doi:10.1038/18613}}.

\bibitem{Osman2021}
A.~Osman, J.~S. Bengtsson, S.~Kose, P.~Krantz, D.~Lozano, M.~Scigliuzzo, P.~Delsing, J.~Bylander, A.~Fadavi~Roudsari, Simplified josephson-junction fabrication process for reproducibly high-performance superconducting qubits, Appl. Phys. Lett. 118 (2021) 064002.
\newblock \href {https://doi.org/10.1063/5.0037093} {\path{doi:10.1063/5.0037093}}.

\bibitem{koch2007}
J.~Koch, T.~M. Yu, J.~Gambetta, A.~A. Houck, D.~I. Schuster, J.~Majer, A.~Blais, M.~H. Devoret, S.~M. Girvin, R.~J. Schoelkopf, Charge-insensitive qubit design derived from the cooper pair box, Phys. Rev. A 76 (2007) 042319.
\newblock \href {https://doi.org/10.1103/PhysRevA.76.042319} {\path{doi:10.1103/PhysRevA.76.042319}}.

\bibitem{Dolan1977}
G.~J. Dolan, Offset masks for lift‐off photoprocessing, Appl. Phys. Lett. 31 (1977) 337.
\newblock \href {https://doi.org/10.1063/1.89690} {\path{doi:10.1063/1.89690}}.

\bibitem{Bilmes_2021}
B.~Alexander, H.~Alexander~K, V.~Serhii, U.~Alexey~V, L.~Jürgen, In-situ bandaged josephson junctions for superconducting quantum processors, Superconductor Science and Technology 34~(12) (2021) 125011.
\newblock \href {https://doi.org/10.1088/1361-6668/ac2a6d} {\path{doi:10.1088/1361-6668/ac2a6d}}.

\bibitem{Slichter}
D.~Slichter, Quantum jumps and measurement backaction in a superconducting qubit, Ph.D. thesis, University of California, Berkeley (2011).

\bibitem{Kreikebaum_2020}
J.~Kreikebaum, K.~O’Brien, A.~Morvan, I.~Siddiqi, Improving wafer-scale josephson junction resistance variation in superconducting quantum coherent circuits, Superconductor Science and Technology 33~(6) (apr 2020).
\newblock \href {https://doi.org/10.1088/1361-6668/ab8617} {\path{doi:10.1088/1361-6668/ab8617}}.

\bibitem{Kirill2011}
K.~Koshelev, M.~A. Mohammad, T.~Fito, K.~Westra, S.~Dew, M.~Stepanova, Comparison between zep and pmma resists for nanoscale electron beam lithography experimentally and by numerical modeling, Journal of Vacuum Science \& Technology B 29~(06F306) (2011).
\newblock \href {https://doi.org/10.1116/1.3640794} {\path{doi:10.1116/1.3640794}}.

\bibitem{Ocola2004}
L.~Ocola, A.~Stein, Effect of cold development on improvement in electron-beam nanopatterning resolution and line roughness, Journal of Vacuum Science \& Technology B 24 (2006) 3061.
\newblock \href {https://doi.org/10.1116/1.2366698} {\path{doi:10.1116/1.2366698}}.

\bibitem{Wenchuang2004}
W.~W. Hu, K.~Sarveswaran, M.~Lieberman, G.~H. Bernstein, Sub-10 nm electron beam lithography using cold development of poly(methylmethacrylate), Journal of Vacuum Science \& Technology B 22 (2004) 1711--1716.
\newblock \href {https://doi.org/10.1116/1.1763897} {\path{doi:10.1116/1.1763897}}.

\bibitem{Cord2006}
B.~Cord, C.~Dames, K.~K. Berggren, J.~Aumentado, Robust shadow-mask evaporation via lithographically controlled undercut, Journal of Vacuum Science \& Technology B: Microelectronics and Nanometer Structures Processing, Measurement, and Phenomena 24~(6) (2006) 3139--3143.
\newblock \href {https://doi.org/10.1116/1.2375090} {\path{doi:10.1116/1.2375090}}.

\bibitem{Tanner2008}
S.~M. Tanner, C.~T. Rogers, Fabrication process for cantilevers with integrated tunnel junctions, Journal of Vacuum Science \& Technology B: Microelectronics and Nanometer Structures Processing, Measurement, and Phenomena 26~(2) (2008) 481--486.
\newblock \href {https://doi.org/10.1116/1.2836428} {\path{doi:10.1116/1.2836428}}.

\bibitem{Potts2001}
A.~Potts, G.~J. Parker, J.~J. Baumberg, P.~de~Groot, Cmos compatible fabrication methods for submicron josephson junction qubits, IEEE Proc.-Sci., Meas. Technol. 148 (2001) 225--228.
\newblock \href {https://doi.org/10.1049/ip-smt:20010395} {\path{doi:10.1049/ip-smt:20010395}}.

\bibitem{SergioO2012}
M.~V. Costache, G.~Bridoux, I.~Neumann, S.~O. Valenzuela, {Lateral metallic devices made by a multiangle shadow evaporation technique}, Journal of Vacuum Science \& Technology B 30~(4) (05 2012).

\bibitem{Lecocq_2011}
L.~Florent, P.~Ioan~M, P.~Zhihui, M.~Iulian, C.~Thierry, F.~Thierry, N.~Cécile, G.~Wiebke, B.~Olivier, Junction fabrication by shadow evaporation without a suspended bridge, Nanotechnology 22~(31) (2011) 315302.

\bibitem{Shankar_2013}
S.~Shankar, M.~Hatridge, Z.~Leghtas, K.~M. Sliwa, A.~Narla, U.~Vool, S.~M. Girvin, L.~Frunzio, M.~Mirrahimi, M.~H. Devoret, Autonomously stabilized entanglement between two superconducting quantum bits, Nature (nov 2013).

\bibitem{Minev_2019}
Z.~Minev, S.~Mundhada, S.~Shankar, P.~Reinhold, R.~Guti{\'{e}}rrez-J{\'{a}}uregui, R.~Schoelkopf, M.~Mirrahimi, H.~Carmichael, M.~Devoret, To catch and reverse a quantum jump mid-flight, Nature (jun 2019).

\bibitem{Pop2012}
I.~M. Pop, T.~Fournier, T.~Crozes, F.~Lecocq, I.~Matei, B.~Pannetier, O.~Buisson, W.~Guichard, Fabrication of stable and reproducible submicron tunnel junctions, Journal of Vacuum Science \& Technology B 30~(1) (2012) 010607.
\newblock \href {https://doi.org/10.1116/1.3673790} {\path{doi:10.1116/1.3673790}}.

\bibitem{LIU2008}
L.~Wei, G.~Zongzhan, Y.~Zhufeng, Steady ratcheting strains accumulation in varying temperature fatigue tests of pmma, Materials Science and Engineering: A 492~(1) (2008) 102--109.
\newblock \href {https://doi.org/10.1016/j.msea.2008.03.042} {\path{doi:10.1016/j.msea.2008.03.042}}.

\bibitem{JeffreyPmma2015}
R.~Jeffrey, J.~Kear, D.~Kasperczyk, X.~Zhang, D.~Chuprakov, R.~Prioul, J.~Schouten, A 2d experimental method with results for hydraulic fractures crossing discontinuities, 2015.

\bibitem{AbdelWAHAB2017}
A.-W. Adel~A., A.~Sabbah, S.~Vadim~V., Temperature-dependent mechanical behaviour of pmma: Experimental analysis and modelling, Polymer Testing 58 (2017) 86--95.
\newblock \href {https://doi.org/10.1016/j.polymertesting.2016.12.016} {\path{doi:10.1016/j.polymertesting.2016.12.016}}.

\bibitem{Devoret1990}
M.~H. Devoret, D.~Esteve, H.~Grabert, G.-L. Ingold, H.~Pothier, C.~Urbina, Effect of the electromagnetic environment on the coulomb blockade in ultrasmall tunnel junctions, Phys. Rev. Lett. 64 (1990) 1824--1827.
\newblock \href {https://doi.org/10.1103/PhysRevLett.64.1824} {\path{doi:10.1103/PhysRevLett.64.1824}}.

\bibitem{Berthold2016}
B.~J\"ack, M.~Eltschka, M.~Assig, M.~Etzkorn, C.~R. Ast, K.~Kern, Critical josephson current in the dynamical coulomb blockade regime, Phys. Rev. B 93 (2016) 020504.
\newblock \href {https://doi.org/10.1103/PhysRevB.93.020504} {\path{doi:10.1103/PhysRevB.93.020504}}.

\bibitem{Baratoff1963}
V.~Ambegaokar, A.~Baratoff, Tunneling between superconductors, Phys. Rev. Lett. 10 (1963) 486--489.
\newblock \href {https://doi.org/10.1103/PhysRevLett.10.486} {\path{doi:10.1103/PhysRevLett.10.486}}.

\bibitem{Macklin}
C.~Macklin, K.~O’Brien, D.~Hover, M.~E. Schwartz, V.~Bolkhovsky, X.~Zhang, W.~D. Oliver, I.~Siddiqi, A near–quantum-limited josephson traveling-wave parametric amplifier, Science 350~(6258) (2015) 307--310.
\newblock \href {https://doi.org/10.1126/science.aaa8525} {\path{doi:10.1126/science.aaa8525}}.

\bibitem{Martinis}
J.~Martinis, Surface loss calculations and design of a superconducting transmon qubit with tapered wiring., npj Quantum Inf. 8~(25) (2022).
\newblock \href {https://doi.org/10.1038/s41534-022-00530-6} {\path{doi:10.1038/s41534-022-00530-6}}.

\end{thebibliography}

\end{document}